# A Model of the Proton

Vladislav Shchegolev

*Flerov Laboratory of Nuclear Reactions, Joint Institute for Nuclear Research, Dubna, Russia*
(Dated: June 10, 2005)

ABSTRACT
I suggest to consider a proton as a body in a state of free precession. Such approach allows to define a proton as periodic system with two rotary degrees of freedom with corresponding frequency ratio and resonances. In result on a power scale the points corresponding to a birth of leptons are designated. The interrelation between masses of leptons is established through a fine structure constant. The given approach is distributed to a nucleus. Other representation is given on the nature of X-rays which connect with a charge of a nucleus and its mass.

This paper proposes a model of the proton that makes possible a new approach to description of elementary physical phenomena and allows one to calculate some fundamental constants[1] previously known only from experiment. The proposed hypothesis is based on the a priori assumption that the proton can be presented as a gyroscope that has a shape of an ellipsoid and undergoes a free precession. Such an approach allows one to describe a proton as the object undergoing simultaneously two rotations, basic and precessional, that creates conditions for appearing resonances between these periodic movements.

## 1. The proton as a gyroscope in a state of free precession

Let us turn to the theory of the gyroscope. This will help us find out why a proton can be thought of as a freely precessing gyroscope and what known physical phenomena can be explained in terms of that concept.

According to the theory of the gyroscope, when a rotor is in a state of steady precession, the OZ axis of the rotor, about which the rotor rotates at a constant angular speed $\varphi$, precesses at a constant angular speed $\Omega$ about the OZ1 axis, which makes an angle $\theta$ with this OZ axis. The components of the external moment **M** for this steady precession, if it has occurred, to be sustained are expressed as

$$M_x = 0$$
$$M_y = [B\varphi + (B - A)\Omega\cos\theta]\Omega\sin\theta$$
$$M_z = 0$$

where **A** - the moment of inertia of ellipsoid with respect to the axis OX, **B** - the moment of inertia of ellipsoid with respect to the axis OZ.

The steady precession can convert into a free precession when the external moment disappears, i.e. at $M_y = 0$, provided $B\varphi + (B - A)\Omega\cos\theta = 0$ or $\Omega = -\dfrac{B\varphi}{(B - A)\cos\theta}$. If A>B and

$\theta$ close to $90^0$, i.e. in the case that it can be taken that $\theta = arctg\dfrac{\Omega}{\varphi}$, we come to expression (1.1).

As long as that condition is fulfilled, the free precession will last indefinitely; at $\theta$ close to $90^0$, the value of $\Omega$ may be very significant[2,3].

$$\frac{1}{\dfrac{A}{B} - 1} = \frac{\Omega}{\varphi}\cos arctg\frac{\Omega}{\varphi} \;, \qquad (1.1)$$

At borderline case when angular velocity of precession $\Omega$ considerably exceeds angular velocity of the basic rotation $\varphi$, axis Z turns on a corner close to $90^0$ (the gyroscope turns sideways, fig. 1b), and as though carries out simultaneously two rotations. Such an extreme case is realized, if **A/B=2+δ** where **δ** has a very small value.

Let us accept *a priori* that the proton is an ellipsoid and, for any reason, is in a free precession state. Thus it makes the basic rotation with angular speed $\varphi$ around the axis OZ and precessional rotation with angular speed $\Omega$ around the axis OX, i.e. the extreme case of the free precession realizes. Its moments of inertia: **B=0.2m$_p$b$^2$** - the moment of inertia of ellipsoid with respect to the axis OZ, **A=0.2m$_p$a$^2$** - the moment of inertia of ellipsoid with respect to the axis OX, (**m$_p$** - mass of a proton, **b** - small half-axle of ellipsoid, **a** - big half-axle of ellipsoid). At **A/B=2+δ** accordingly $a/b = \sqrt{2 + \delta}$.

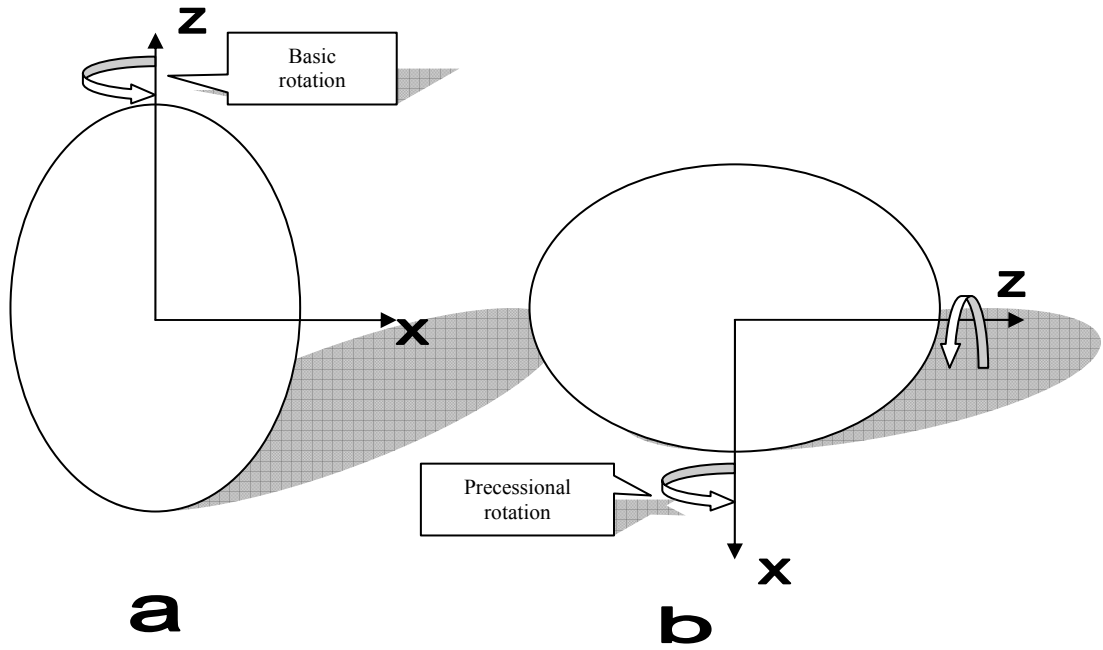

Fig.1  a) Basic rotation with respect to the axis OZ.
   b) Free precession with respect to the axis OX.

We allow that in the course of its appearing in the Big Bang the proton has come in a state of the free precession. Thus it has incorporated a maximum of the energy concentrated in its precessional rotation due to what linear speed of the ellipsoid's top has achieved the greatest possible speed – the speed of light. How will it react to any external influence? It should tear away the energy brought from the outside. According to the third law of Newton, the gyroscope develops equal on value and the opposite moment on a direction, interfering external influence. This moment M is called a gyroscopic reactionary torque. But this reaction should be expressed in a specific way. As the proton in our representation is a periodic system with two rotary degrees of freedom, the fault of energy should be carried out according to the frequency correlations inherent in periodic systems and resonances. In other words, the excessive energy should be removed by electromagnetic quantums or irraradiation of particles. The latter is the assumption that should be checked-up.

## 2. Proton and radiations

If our proposed idea takes place indeed, then it is possible to assume that appearance of the electron is a result of some external influence on proton being in a free precession state.

Being guided by only intuitive guess, we take that at the moment of electron appearance the relation of angular velocity of precession to angular speed of the basic rotation is equal to $\Omega/\varphi$ = 1836.156, i.e. to the relation of the proton mass to electron mass $m_p/m_e$. Then according to (1.1) $A/B$=2.0000082 and, accordingly, it is possible to take $a/b=\sqrt{2}$.

As we assumed in Section 1, in the considered extreme case the linear speed of the top of the ellipsoid-proton making precessional rotation, is equal to the speed of light $c$=2.997924·$10^8$m/s, i.e. $V_{pr}=\Omega a=c$. As frequencies of the basic and precessional rotations of the gyroscope are interconnected, we can determine the linear speed of the end of small half-axle $V_{main}$ which with the account $a/b=\sqrt{2}$ and $\Omega/\varphi$ = 1836.156 is calculated from (2.1):

$$V_{main} = \varphi b = \frac{1}{\sqrt{2}} \frac{\varphi}{\Omega} V_{pr} = 1,15450 \cdot 10^5 m/s \qquad (2.1)$$

Taking into account that the moment of inertia equals to $B = \frac{1}{5}mb^2$, the kinetic energy of basic (main) rotation $E_{main}$ is equal to:

$$E_{main} = \frac{B\varphi^2}{2} = \frac{0.2 m_p b^2}{2}\left(\frac{V_{main}}{b}\right)^2 = 0.1 m_p V_{main}^2 = 13.914928 eV \qquad (2.2)$$

The value $E_{main}$ with accuracy of 2.3 % coincides with Ridberg energy $E_R$ =13.605667 eV which is determined by the top border of Layman series (the so-called optical limit) and is taken as the energy of ionization of the hydrogen atom. This result is a surprising one since it is obtained on the basis of only mechanistic representation and provides a way to consider the suggested model in more detail. Therefore, we shall continue our estimations. To the explanation of the difference between $E_{main}$ and $E_R$ we shall return later.

*Interrelationship of fundamental constants in frames of the model*

If we knew a numerical value of $\nu_{main}$, the frequency of the basic rotation, we would calculate the value **b** under the formula

$$b = \frac{V_{main}}{\varphi} = \frac{V_{main}}{2\pi\nu_{main}}. \qquad (2.3)$$

Let us make an arbitrary assumption that $\nu_{main}=c/\Lambda_e$ ($\Lambda_e$=2.426310·10⁻¹²m – Compton length of electron) that gives value $\nu_{main}$=1.23559·10²⁰ Hz. Validity of this assumption is justified by the results obtained below. We transform, accordingly, the formula (2.3) into (2.4):

$$b = \frac{V_{main}}{2\pi\nu_{main}} = \frac{V_{main}\Lambda_e}{2\pi c}, \qquad (2.4)$$

where $V_{main}$= 115450.76m/s from (2.1). Calculation by formula (2.4) gives **b**=1.48711·10⁻¹⁶m and, accordingly, $a=\sqrt{2}$ **b**=2.103090·10⁻¹⁶m, that is in agreement with the known experimental data about the proton sizes.

It is uneasy to be convinced that the equality is precisely held:

$$2\pi a = \Lambda_p, \qquad (2.5)$$

where $\Lambda_p$ is a length of the circumference described by radius **a**, and precisely coincides with the Compton length of proton $\Lambda_p$=1.321409·10⁻¹⁵m.

Also it is possible to be convinced that the equality is so precisely held:

$$m_p c a = 1.054571481 \cdot 10^{-34} Js = \frac{h}{2\pi} \qquad (2.6)$$

i.e. the angular momentum of a point with mass $m_p$ moving on a circumference of radius **a**, is equal to Planck's constant **h** divided on **2π** (ℏ=h/2π).

Thus, in the framework of the proposed model and accepted arbitrary $\nu_{main}=c/\Lambda_e$ we have precisely calculated the sizes of the proton, the Compton length of the proton and Planck's constant. Hence, it is possible to approve that the assumption is correct.

Let us comment the obtained result. The concept of the Compton length of electron $\Lambda_e=h/m_e c$ appeared in 1923 after the experiment performed by Compton on the angular scattering of X-rays. It has been shown that the spectra of scattered X-rays displace aside longer waves in comparison with the spectra of initial X-rays, and the value of displacement Δλ depends only on scattering angle θ. The experimental data were fitted by the formula $\Delta\lambda = C\sin^2\frac{\theta}{2}$, where **C** was the universal constant having a length dimension and numerically equal to $2\frac{h}{m_e c}$. At theoretical interpretation of the Compton effect, the following physical assumption has been entered: if the X-ray quantum scattered by electron in atom and takes up a part of energy and a impulse of the falling quantum. The laws of conservation of energy and a impulse were applied. The quantum energy was determined as $E_\gamma=h\nu$, the impulse as $p_\gamma=h\nu/c$ (by A.Einstein). Thus the formula $\Delta\lambda = 2\dfrac{h}{m_e c}\sin^2\dfrac{\theta}{2}$ has been obtained. But it is necessary to note that occurrence of universal value $\Lambda_e=h/m_e c$ in this formula was only formal, following mathematical calculations. Reflections of physicists about physical sense of $\Lambda_e$ have led to representation about the dual nature of radiation – wave-corpuscular. According to this conception, the particle can be presented as a quantum with energy $E_\gamma=h\nu=mc^2$ and, accordingly, as a wave with a wavelength $\Lambda=c/\nu=h/mc$.[4] The Compton-effect has been recognized as the experimental evidence of this dual nature. In our representation the Compton length has another physical maintenance. With reference to proton $\Lambda_p$ corresponds to frequency of free precession. With reference to electron $\Lambda_e$ corresponds to frequency of the basic rotation of the proton at the moment of electron birth. How to interpret the Compton effect in this representation? The author is at a loss to answer. For this purpose, one should assume that there is a certain interaction between the frequency of X-ray radiation and the frequency of the basic rotation of the proton.

From the previous consideration it is possible to deduce more conclusions. Let us calculate the frequency of precession $\nu_{pr}$ from $c=\Omega a=2\pi\nu_{pr}a$: $\nu_{pr}=2.26873\cdot 10^{23}$ Hz. The energy of the quantum corresponding to this frequency of rotation: $E=h\nu_{pr}=938.27157$ MeV, that coincides with the energy of the proton rest mass, calculated under Einstein's formula $E=m_p c^2$.

Let us see that the energy determined under formula $E=m_p c^2$ is the total energy of a thin ring with mass $m_p$ and radius $a$ rotating with angular velocity $\omega_{pr}$.

$$E = I\omega_{pr}^2 = (m_p a^2)(2\pi\nu_{pr})^2 = (4\pi^2 m_p a^2 \nu_{pr})\nu_{pr} = h\nu_{pr} \qquad (2.6a)$$

It is easy to be convinced that with the calculated values $a$ and $\nu_{pr}$ product $4\pi^2 m_p a^2 \nu_{pr}=h$. Introduction of analogy between the precessional rotation of a proton-ellipsoid and the rotation of the thin ring, having the same mass and frequency of rotation, has allowed us to look differently at the nature of the energy of rest mass.

However, there is a question. The energy of any quantum is determined as $E_\gamma=h\nu$, where $\nu$ - the frequency inherent in quantum. At the same time, the earlier calculated value $h$ includes as a multiplier frequency $\nu_{pr}$, inherent to precessional rotation of the proton. It turns out that $E_\gamma$ is proportional to product $\nu_{pr}\nu$. According to the theory of a gyroscope in precession state, this should be. The matter is that the reactive moment M with which a gyroscope reacts to external influence, is determined as $M=I\omega_{pr}\omega_{main}$, where $I$ - the moment of inertia, $\omega_{pr}$ - angular speed of precession, $\omega_{main}$ - angular speed of the basic rotation[3]. According to Euler's equation for a rotating body in a state of free precession angular speed $\omega_{pr}$=const[5] From here it is clear why $h$=const. Clearly also, the reaction of such a gyroscope on the external influence should be expressed in change of angular speed of the basic rotation and to corresponding dump of energy as radiation of particles or $\gamma$-quanta. It is shown below how it occurs in reality.

The proposed model allows one to obtain some more interesting results. We shall use an analogy between the rotating ring and the quantum having the same frequency.

Let us return to the formula (2.3) and make one more assumption. We admit that the energy of basic rotation $E_{main}=13.914928$ eV calculated by (2.2) is the energy of the quantum having a corresponding frequency, according to the formula of Planck $E_{main}=h\nu$. On the other hand, we shall define the same energy as energy of a rotating ring with mass $m_p$ and with the frequency equal to the frequency of quantum. For a rotating ring kinetic energy is determined as

$$E_{main} = \frac{1}{2}I\omega^2 = \frac{1}{2}m_p b_r^2 \left(\frac{V_r}{b_r}\right)^2 = \frac{1}{2}m_p V_r^2, \qquad (2.7)$$

where $I = m_p b_r^2$ - ring moment of inertia, $b_r$ – radius, $V_r$ - linear velocity. Let us define $b_r$ as

$$b_r = \frac{V_r}{2\pi\nu} = \frac{h}{2\pi\nu h}\sqrt{\frac{2E_{main}}{m_p}} = \frac{h}{2\pi}\sqrt{\frac{2}{E_{main}m_p}} = 2.442287 \cdot 10^{-12} \text{m} \qquad (2.8)$$

It turns out that value $b_r$ almost coincides with the value of Compton length for electron $\Lambda_e = h/m_e c = 2.426310 \cdot 10^{-12}$m and ratio $k_1 = \dfrac{b_r}{\Lambda_e} = 1.006584242 = \dfrac{\sqrt{10}}{\pi}$. The physical sense of factor $\kappa_1$ will be explained below. What can this result mean?

It is possible to conclude that we deal with the latent parametrical resonance at which the wave length $\Lambda_e$ coincides with the radius of the ring determined above, rotating with frequency $\nu = E_{main}/h$. The discrepancy of value $b_r$ and $\Lambda_e$ on the factor $\kappa_1$ confirms this conclusion, and that is why.

In the theory of resonance it is shown that the resonance in an oscillatory system should arise at stimulating frequency $\nu$ which is smaller than resonant frequency $\nu_0$[6]. A ratio between $\nu$ and $\nu_0$: $\dfrac{\nu}{\nu_0} = 1 \pm \dfrac{1}{2Q}$, where Q – tuned-circuit Q-factor, the value that determines power losses in an oscillatory circuit. To compensate these losses and to reach resonant frequency, an additional energy should be brought in the resounding system. Indeed, we see that at energy $E_{main}=13.914928$ eV the frequency corresponding $\Lambda=2.442287 \cdot 10^{-12}$m is less than the resonant frequency corresponding $\Lambda_e$. Let us determine what should be energy $E_{main}$ at which the resonant frequency has been achieved.

To make this, we transform (2.8) into (2.9):

$$\frac{h}{m_e c} = \frac{h}{2\pi}\sqrt{\frac{2}{E_{main}m_p}} \qquad (2.9)$$

After simple algebraic manipulations we receive

$$(m_e c^2)^2 = 2\pi^2 E_{main} m_p c^2. \qquad (2.10)$$

Having substituted in (2.10) the numerical values of electron and proton masses, we received $E_{main}=14.098727$ eV whose exceeding value obtained earlier on ~0.2 eV.

Let us find out on what this additional energy ~0.2 eV is necessary. We shall make one more transformation of expression (2.9) substituted in it $E_{main} = 0.1 m_p V_{main}^2$ according to (2.2).

$$V_{main} = \frac{\sqrt{10}}{\pi} \cdot \frac{c}{\dfrac{m_p}{m_e}\sqrt{2}} = 116210.9173 \frac{m}{s} \qquad (2.11)$$

Formula (2.11) expresses in essence the conservation law of an impulse. Clearly, the additional energy ~0.2 eV is spent for feedback of the proton at an electron birth. From (2.11) it is seen also how factor $k_1 = \dfrac{\sqrt{10}}{\pi}$ appears.

Here the following should be clarified. The identification of the rotating body with a rotating thin ring (it is equivalent to a point moving in a circle) with the same mass, frequency of rotation and, accordingly, energy of rotation in the same measure is proved, as well as an identification of body forward motion with motion of its center of mass. However, when observing the energy equality, the linear orbital speeds differ from each other.

$$\frac{1}{10}m_p V_{elip}^2 \cdot \frac{\pi^2}{10} = \frac{1}{2}m_p V_{circl}^2; \quad \frac{V_{elip}}{V_{circl}} = \sqrt{\frac{50}{\pi^2}} = 2,25 \tag{2.11a}$$

*Hypothetical particles*

It is necessary to elucidate one interesting question: what mean Rydberg energy $E_R$ in our consideration? In Bohr`s model of hydrogen atom $E_R$ is electron kinetic energy in 1-st orbit and is accepted as upper limit of an optical spectrum. Under formula $E_R=h\nu$ the value $\nu=3.28987 \cdot 10^{15}$ Hz coincide fine with upper limit of short-wave Layman series. This fact has defined success of the planetary model. In our model energy $E_{main}=14.098727$ eV is close to $E_R$ (a difference ~0,5 eV). But we shall show that $E_R$ has other physical contents.

It was shown above that arbitrary choice of the basic rotation frequency $\nu_{main}=c/\Lambda_e=1,23 \cdot 10^{20}$ Hz appeared to be proved due to the fact of coincidence of values $b_r$ and $\Lambda_e$ (a condition of resonance). It has allowed one to calculate values of some fundamental constants correctly. But this value $\nu_{main}$ corresponds to the upper limit of X-ray range on a power scale. In the logic of our consideration it turns out that the electron birth occurs at this frequency. Then due to the same logic, we should assume that at frequency $\nu=3.2880 \cdot 10^{15}$ Hz calculated under formula $E_R=h\nu$ there is a birth of particle X with mass $m_x c^2 = E_R = 13,605$ eV. We shall show that is so. According to (2.1), at birth of this particle $V_{main} = c \frac{E_R}{m_p c^2 \sqrt{2}} = 3,0723$ m/s. Having substituted this value $V_{main}$ into (2.3) with $b=1.48711 \cdot 10^{-16}$ m, we obtain $\nu_{main} = \frac{V_{main}}{2\pi b} = 3,2880 \cdot 10^{15}$ Hz, as it was required to show.

But there is a question: what does the particle with mass $m_x c^2 = E_R = 13,605$ eV mean? If it really exists, we should assume that it have an electric charge. It is possible to explain occurrence of charges at triboelectrification or in the piezoelectric phenomena because the energy necessary for such excitation is small enough ($V_{main} \approx 3$ m/s) Secondly, its mass is close to the expected neutrino mass, and on this basis it is possible to name this particle "electrino". The following question: how to compare it to the available representations about hydrogen atom? We shall try to answer this question below.

The resonant approach suggested here allows one to solve another problem: the problem of definition of the muon mass. It is clear that the ratio

$$\frac{\Lambda_p}{b} = \frac{1.321409 \cdot 10^{-15} м}{1.48711 \cdot 10^{-16} м} = 8.88575, \tag{2.12}$$

practically coincides with the ratio of the proton mass to muon mass $\frac{m_p}{m_\mu} = 8.88024$. It gives a chance to assume that the muon birth corresponds to this resonance. The calculations by formulas (2.11), (2.2) and (2.10) at condition $\frac{\Omega}{\varphi} = \frac{\Lambda_p}{b} = 8.88575$ give the following result:

$$V_{main} = \frac{\sqrt{10}}{\pi} \cdot \frac{c}{\frac{\Lambda_p}{b}\sqrt{2}} = 2.401384 \cdot 10^7 \frac{m}{s}$$

$$E_{main} = 0.1 m_p c^2 \frac{V_{main}^2}{c^2} = 0.60202 MeV$$

$$m_x c^2 = \sqrt{2\pi^2 E_{main} m_p c^2} = 105.59286 MeV$$

The calculated value $m_x c^2$ differs from the experimental value of muon mass $m_\mu c^2$= 105.65839 MeV by the factor of 1.0006.

Thus, a row of leptons is built: muon, electron and hypothetical electrino whose masses are defined equally and correspond to the certain resonances.

There is a question: whether is it possible to connect the masses of these particles by a uniform dependence? It is guessed in expression (2.10). In case of a muon birth $E_{main}$=0.602 MeV and differs from a rest mass energy of electron $m_e c^2$=0.511 meV by the factor of 1.178. In case of electron birth $E_{main}$=14.1 eV it differs from the rest mass energy of electrino $m_{el} c^2$=13.6 eV already by the factor of 1.036. It would be seductive to use an outlined law for extrapolations aside low and high energies. Actually, the question on interrelation of leptons masses is not so simple. We shall consider it in Section 5. Here we are limited to the assumption that expression (2.10) can be used for calculation of the critical energy necessary for electrino birth, i.e.

$$E_{main}^{el} = \frac{(m_{el} c^2)^2}{2\pi^2 m_p c^2} = 0.9995 \cdot 10^{-8} eV \approx 1 \cdot 10^{-8} eV$$

While we shall be limited to this assumption, however now there is an idea that on a power scale there are critical values of energy at which there is a birth of corresponding particles in a resonant process. That is a reaction of proton in a free precession state. This brings up the question: what will take place if the energy of external influence gets in an interval between the specified critical values? The answer is prompted by experience. The fault of excessive energy will be carried out by electromagnetic radiation, but according to frequency parities, installation-specific for periodic systems. We shall demonstrate the correctness of this statement by an example of optical spectra.

In our presentation the resonant condition can be written as follows:

$$(m_x c^2)^2 = 2\pi^2 E_{main}^x m_p c^2 \tag{2.13}$$

As we found out above, we can replace value $m_x c^2$ on $h\nu_x$ and rewrite (2.13) as follows:

$$(h\nu_x)^2 = 2\pi^2 \kappa m_{x-1} c^2 m_p c^2 \tag{2.14}$$

***Optical spectrum of atomic hydrogen***

We shall be convinced below that the optical spectrum of the hydrogen atom is formed by frequencies $\nu^*$ in a range from $\nu = \frac{E_{main}^{el}}{h}$ up to $\nu = \frac{m_{el} c^2}{h}$ determined by expression (2.15):

$$(n^2 h\nu^*)^2 = 2\pi^2 E_{main}^{el} m_p c^2, \tag{2.15}$$

where **n** – integer, $\nu^*$ - frequency of corresponding overtone. The frequency of optical waves in Layman series is determined as

$$\nu_L = \nu_{el} - \nu^* \tag{2.16}$$

Results of calculations are tabulated (see Table 1). The good consent between calculated and experimental data is visible.

Table 1

The energies of optical quanta of Layman series

| n | $h\nu^*$ eV | $h\nu_L$ eV | $\lambda = c/\nu_L$ nm | $\lambda_{exp}$ nm |
|---|---|---|---|---|
| 2 | 3.401 | 10.204 | 121.50 | 121.57 |
| 3 | 1.511 | 12.094 | 102.52 | 102.60 |
| 4 | 0.850 | 12.808 | 96.80 | 97.27 |
| 5 | 0.544 | 13.061 | 94.92 | 94.97 |

Calculation of wavelengths of other series is carried out under the following formulas. The Balmer series: $h\nu_L = h\left(\nu^*_{n=2} - \nu^*_{n=3,4,5...}\right)$. The Paschen series: $h\nu_L = h\left(\nu^*_{n=3} - \nu^*_{n=4,5,6...}\right)$. The Brackett series: $h\nu_L = h\left(\nu^*_{n=4} - \nu^*_{n=5,6,7...}\right)$. Results are shown in Table 2.

Table 2

| The Balmer series | | | The Paschen series | | | The Brackett series | | |
|---|---|---|---|---|---|---|---|---|
| $h\nu_L$ eV | $\lambda = c/\nu_L$ nm | $\lambda_{exp}$ nm | $h\nu_L$ eV | $\lambda = c/\nu_L$ nm | $\lambda_{exp}$ nm | $h\nu_L$ eV | $\lambda = c/\nu_L$ nm | $\lambda_{exp}$ nm |
| 1.889 | 656.14 | 656.28 | 0.661 | 1874.6 | 1875.1 | 0.306 | 4050.4 | 4050 |
| 2.551 | 486.02 | 486.13 | 0.967 | 1281.5 | 1281.7 | 0.472 | 2624.5 | 2630 |
| 2.857 | 433.93 | 434.05 | 1.133 | 1093.5 | 1090 | 0.5725 | 2164.9 | - |
| 3.023 | 410.07 | 410.17 | 1.2340 | 1004.7 | 1004.9 | 0.6375 | 1944.2 | - |
| 3.123 | 396.91 | 397.01 | 1.2991 | 954.4 | 954.6 | | | |
| 3.188 | 388.81 | 388.90 | | | | | | |

Thus, in the proposed model the radiation is defined by differences of frequencies of the proton basic rotation. The choice of frequencies $\nu^*$ submits to the rule $n^2 h\nu^* = h\nu_x = m_x c^2$. The particle birth occurs at the frequencies corresponding to a condition of resonance (2.13). This concept differs from Bohr's model according to which radiation takes place between the quantized energy levels of electron rotating around the proton. The proposed model removes the difficulties of classical physics which have arisen in due time at interpretation of the optical spectrum of the hydrogen atom. Moreover, the model does not require use of electric interaction between proton and electron. We shall say about the nature of the electric interaction with reference to the nuclear phenomena below.

### 3. Spin

Let us transform expression (2.9) $\dfrac{h}{m_e c} = \dfrac{h}{2\pi}\sqrt{\dfrac{2}{E_{main} m_p}}$ with the account that in the framework of the analogy of rotating ring $E_{main} = \dfrac{1}{2} m_p V_r^2$ ($V_r$-linear speed of the ring). After simple transformations we receive

$$m_e \frac{c}{\pi} = m_p V_r \qquad (3.1)$$

The formula (3.1) expresses the conservation law of the impulse. We obtain an interesting result. After birth, the electron has speed $c/\pi$. The kinetic energy corresponding to this speed we attribute to the rotary energy (spin energy). It is equal to

$$E^{kin}_{rot} = \frac{m_e c^2}{2\pi^2} = 25887.5 эB \approx 0.026 MэB. \qquad (3.2)$$

According to suggested model primary presence of electron in hydrogen atom not necessarily. Its occurrence is connected with external influence of resonant character. To reduce power balance, it is necessary to assume, that the mass of hydrogen atom (protium) in power units at least is equal: $M_{prot} c^2 = m_p c^2 + m_e c^2 + \dfrac{m_e c^2}{2\pi^2}$ =938,8087 МэВ (1,0078526 а.е.м.). The accepted mass value of protium is equal to 1,007825 amu[7]. (1 amu = 1,6605388·10⁻²⁷кг) Difference in 1,000027 amu is not dramatic. It is stacked in uncertainty of atomic mass of hydrogen (from 1,0077 amu up to 1,0081 amu)[7].

It will be shown below that introduction of spin energy of electron in model is rather essential circumstance in definition of an elementary electric charge, fine structure constant and masses of leptons.

## 4. About the nature of elementary electric charge

Let us build the following succession of reasonings. In the framework of the analogy of a material point rotating in a circle (or ring), the centripetal force F acting on a point is equal to $F = \dfrac{mV^2}{r}$ (m - weight, V - orbital velocity, r - radius of a circle). We shall redefine force F through full energy E and the angular moment L as follows:

$$F = \frac{mV^2}{r} \cdot \frac{mV^2}{mV^2} = \frac{(mV^2)^2}{(mVr)V} = \frac{E^2}{LV} \quad \text{or} \quad LV = \frac{E^2}{F} = const \qquad (4.1)$$

The product LV has dimension $\left[ kg \cdot \dfrac{m}{s} \cdot m \cdot \dfrac{m}{s} \right] = kg \cdot \dfrac{m^3}{s^2}$. Dimension $\dfrac{m^3}{s^2}$ corresponds to value $r^3\nu^2$ ($\nu$ – a rotation frequency). In accordance with the third Keplerian law for rotating bodies $r^3\nu^2 = const$.

We have established in Section 3 the energy $E = m_e c^2 + \dfrac{m_e c^2}{2\pi^2}$ is need at least for the electron birth. It realizes in the rotary movement that we associate with the motion of a point in a circle. Accordingly, we can accept E=**h**$\nu$. For a rotating point the orbital velocity is V=**2π$\nu$r**. We define a radius of such rotations (back) as $r = \dfrac{h}{2\pi^2 m_e c}$. Then $V = \dfrac{h\nu}{\pi m_e c}$. Thus, the velocity V is equal to

$$V = 2\pi\nu \frac{h}{2\pi^2 m_e c} = \frac{h\nu}{\pi m_e c} = \frac{m_e c^2 + \dfrac{m_e c^2}{2\pi^2}}{\pi m_e c} = \frac{c}{\pi}\left(1 + \frac{1}{2\pi^2}\right) \approx \frac{c}{3} \qquad (4.2)$$

Then in accordance with (4.1)

$$F = \frac{mV^2}{r} = m_e \frac{2\pi^2 m_e c}{h} \cdot \frac{c^2}{\pi^2}\left(1 + \frac{1}{2\pi^2}\right)^2 \cdot \frac{c}{c} = \frac{2(m_e c^2)^2}{hc}\left(1 + \frac{1}{2\pi^2}\right)^2 \qquad (4.3)$$

On the other hand, the kinetic energy of spin rotation is equal to $E_{sp}^{kin} = \dfrac{m_e c^2}{2\pi^2}$, because V$_e$=**c/π** according to (3.1). Hence, according to (4.1) for that rotation

$$(LV)_{sp} = \frac{1}{F} \cdot \left(\frac{m_e c^2}{2\pi^2}\right)^2 \qquad (4.4)$$

If to accept that force F is the same as in (4.3) and in (4.4), then we obtain

$$(LV)_{sp} = \frac{hc}{8\pi^4\left(1 + \dfrac{1}{2\pi^2}\right)^2} = const \qquad (4.5)$$

Let us present dimension LV as $\left[ kg \cdot \dfrac{m}{s} \cdot m \cdot \dfrac{m}{s} \right] = \left[ \dfrac{kg \cdot m}{s^2} \cdot m^2 \right] = [n \cdot m^2]$ and express (LV)$_{sp}$ as (LV)$_{sp}$=**fr²**. Being guided by an intuitive guess, we take **f** as a Coulomb force. Then $(LV)_{sp} = \dfrac{q^2}{4\pi\varepsilon_0 r^2} \cdot r^2$. After simple operations expression (4.5) takes the form:

$$q^2 = \frac{\varepsilon_0 hc}{2\pi^3 \left(1 + \frac{1}{2\pi^2}\right)^2} = \frac{2\pi\varepsilon_0 hc}{(2\pi^2 + 1)^2} \qquad (4.6)$$

The calculated value **q**=1,6029·10$^{-19}$ C differs from experimental value **e**=1,602176·10$^{-19}$ C on factor 1,00046.

It is noteworthy that mass of the particle was excluded when we derived the formula (4.5). It means finally that any irradiated particle should have the same electric charge. And naturally one can make a conclusion that the constancy of the elementary electric charge is a consequence of the 3-rd law of Kepler, and its absolute value is initially set by velocity **c/π** at which a birth of any particle takes place.

The obtained result brings us to a fundamental question on a physical sense of the elementary electric charge. In the book "Electrodynamics" [8] A.Sommerfeld writes that we are compelled to accept existence of electric charges as a reality of the Nature. All attempts to reduce the electric phenomena to mechanical ones suffered failure. The charge was prescribed own dimension, exceeding the bounds of mechanics. Maxwell`s theory does not consider the atomic structure of the charge at all. The charge is taken as an absolute constant. In Maxwell`s theory the concept of vector **D**, dielectric displacement (or electric excitation under offer Sommerfeld) is entered. To connect the concept of the charge with the viewpoint of the field theory, it is necessary to imagine some excitations of the environment which start with the charged center. The concept of the charge is connected historically to the theory of the long-range action that existed in the 2-nd half of the XIX century and was practically forgotten. This theory explained the phenomena of attraction and pushing away as the interaction of periodic systems[9].

Return to this theory can give a key to understanding the result obtained above with calculation of the elementary electric charge value because it was obtained owing to consideration of birth of the charged particle as a resonance in a periodic process. After birth, the proton and electron, being in turn periodic systems, interact with each other, probably, under the laws established by the theory of long-range action. This question is quite complicated and demands special studying.

### 5. A fine structure constant and interrelations between lepton masses.

The fine structure constant $\alpha = \dfrac{e^2}{2\varepsilon hc} = \dfrac{1}{137.036}$ (in system of SI) plays an important role in modern physics. Its occurrence in atomic physics is connected to observed splitting of spectral lines of atoms explained by spin-orbital interaction. All distances between the components of splitting a line can be expressed as a product of **α²** on rational numbers. The theory of fine structure is built on the basis of relativistic consideration of keplerian motion of electron around the nucleus. In quantum electrodynamics the value **α** is taken as a constant of interaction of photon with electron. A. Sommerfeld emphasized a special role of this constant because in it three mainstreams of modern physics were reflected: the electronic theory (a charge **e**), the quantum theory (Planck's constant **h**) and the theory of relativity (speed of light **c**).

The fine structure constant appears in our consideration as follows:

$$\frac{m_e c^2}{\pi^2 m_p c^2} \cdot \frac{E_R}{E_{main}} = \alpha^2 = \frac{1}{(137,036)^2} \qquad (5.1)$$

i.e. as a relation of total spin energy $\dfrac{m_e c^2}{\pi^2}$ of electron and the rest mass of proton multiplied by the relation of $E_R$=13,605667 eV and **E$_{main}$**=14,098727 eV (see 2.10).

The equality (5.1) should hold for all elementary charged particles because **α** is a universal constant. We shall see how it holds in a line of leptons if the particle with the rest mass $m_{x-1}c^2$ is a "predecessor" of the particle with the rest mass $m_x c^2$. We designate the threshold energy at which the particle with mass $m_x c^2$ is born as $E_{cr}^x$. In case of electron birth $m_{x-1}c^2 = m_{el}c^2 = 13,605667$ eV and $E_{cr} = E_{main} = 14,098727$ eV. The generalized formula is the following:

$$\frac{m_x c^2}{\pi^2 m_p c^2} \cdot \frac{m_{x-1} c^2}{E_{cr}^x} = \alpha^2. \tag{5.2}$$

If to start from **τ**-particle ($m_\tau c^2 = 1777$ MeV), then for it $E_{cr}^\tau = \frac{m_\tau c^2}{\pi^2 m_p c^2} \cdot \frac{m_\mu c^2}{\alpha^2} = 3,8 \cdot 10^5$ MeV. The following chain can be obtained:

$$E_{cr}^\tau = 3,8073 \cdot 10^5 Mev = 214,2545 m_\tau c^2 = m_\tau c^2 \frac{m_\mu c^2}{m_e c^2} \cdot 1,03624 = m_\tau c^2 \frac{m_\mu c^2}{m_e c^2} \cdot \frac{E_{cr}^e}{m_{el} c^2}.$$

For muon:

$$E_{cr}^\mu = \frac{m_\mu c^2}{\pi^2 m_p c^2} \cdot \frac{m_e c^2}{\alpha^2} = 109,4878 \text{ MeV} = 1,03624 m_\mu c^2 = m_\mu c^2 \frac{E_{cr}^e}{m_{el} c^2}$$

For the sake of generality we introduce a W-boson in the considered row of particles ($m_W c^2 = 80,33$ GeV). As a result of the performed calculations, we get exact ratios:

$$\frac{E_{cr}^W}{E_{cr}^\tau} = \frac{m_W c^2}{m_\mu c^2}; \quad \frac{E_{cr}^\tau}{E_{cr}^\mu} = \frac{m_\tau c^2}{m_e c^2}; \quad \frac{E_{cr}^\mu}{E_{cr}^e} = \frac{m_\mu c^2}{m_{el} c^2} \tag{5.3}$$

$$(m_W c^2)^2 = \frac{(m_\tau c^2)(m_\mu c^2)(m_e c^2)}{(1,03)^3 (m_{el} c^2)} \tag{5.4}$$

Formula (5.4) gives a W-boson mass with accuracy of Δ=1,000027.

Thus, the interrelation of lepton masses with each other and with the proton mass is established through a fine structure constant. The calculated threshold energies coincide with experimental ones.

It is seen that a discrepancy arises with threshold value $E_{cr}^e = 14,1$ eV which appeared to be 3,6 % more than the taken value of potential of ionization of hydrogen atom of hydrogen $E_{ion}^H = 13,6$ eV. But it should be noted that this value was not obtained in direct experiment. Sommerfeld took note of that [4]. Different values were obtained at measurement of $E_{ion}^H$ by various methods. For example, $E_{ion}^H = 15,34$ eV was measured in experiences with electronic impacts in system $H_2 \rightarrow {}^+H_2$+electron. Finally, it was agreed to take $E_{ion}^H$ on the top border of Layman series (on an optical limit). In our consideration $E_{ion}^H = 14,098727$ eV, and $E_R = 13,605667$ eV is rest mass $m_{el}c^2$ of the particle named "electrino".

Now we return to the question on possible existence of younger leptons. In Section 2 we determined the critical energy necessary for electrino birth as $E_{cr}^{el} = \frac{(m_{el} c^2)^2}{2\pi^2 m_p c^2} = 0,99949627 \cdot 10^{-8} eV$. Having substituted this value in (5.2), we received a rest mass of the particle, the predecessor of electrino which we name a **ch**-particle:

$$m_{ch} c^2 = \frac{\pi^2 \alpha^2 m_p c^2 E_{cr}^{el}}{m_{el} c^2} = 3,6226 \cdot 10^{-4} eV \tag{5.5}$$

It is uneasy to be convinced that the relation of kinetic spin energy of a **ch**-particle $E_{sp}^{ch} = \frac{m_{ch} c^2}{2\pi^2}$ to $E_{cr}^{el}$ precisely is equal to the relation of masses of proton and electron:

$$\frac{m_{ch}c^2}{2\pi^2 E_{cr}^{el}} = \frac{\alpha^2 m_p c^2}{2m_{el}c^2} = 1836,156 = \frac{m_p c^2}{m_e c^2} \tag{5.6}$$

Reducing the second part of equality (5.6) to $m_p c^2$, we receive

$$2\frac{m_{el}c^2}{m_e c^2} = \alpha^2 \tag{5.7}$$

Thus we have received a very simple ratio for the masses of electrino and electron. From the received dependences it is possible to obtain some other simple ratios:

$$\left(m_{el}c^2\right)^2 = \left(m_{ch}c^2\right)\left(m_e c^2\right) \tag{5.8}$$

In a general view:

$$2\frac{m_{x-1}c^2}{m_x c^2} = \alpha^2; \quad \left(m_x c^2\right)^2 = \left(m_{x-1}c^2\right)\left(m_{x+1}c^2\right); \quad m_{x-1}c^2 = \frac{\alpha^4}{4} m_{x+1}c^2 \tag{5.9}$$

Those ratios are performed exactly in row: electron – electrino – **ch**-particle.

How much do these hypothetical particles correspond to the reality? We shall show that the ch-particle concerns a fine structure of lines of the optical spectrum of atomic hydrogen. Experiment[10] has shown that in Balmer series lines are split into doublets with a constant distance between components $\Delta v \approx 36,45 м^{-1}$ (in a scale of wave numbers). In Sommerfeld`s theory which was developed on the basis of a relativistic consideration of keplerian motion of electron around the nucleus, the value $\Delta v$ is expressed as $\Delta v_H = \frac{1}{2^4} R\alpha^2 = 36,523 m^{-1}$ (Rydberg constant R=1,0973742·10⁷m⁻¹).

In our consideration this value is expressed with account of (5.8) as

$$\Delta v = \frac{R}{2^3} \frac{m_{ch}c^2}{m_{el}c^2} \tag{5.10}$$

On the other hand, it is 8 times higher the wavelength (8 times less than frequency) corresponding to the ch-particle:

$$\frac{1}{\Delta v} = 2,738 \cdot 10^{-2} м; \quad \lambda_{ch} = \frac{hc}{m_{ch}c^2} = 3,42252 \cdot 10^{-3} м; \quad \frac{1}{\Delta v} = 8\lambda_{ch} \tag{5.11}$$

The observed wave shift corresponds to $\frac{1}{8}m_{ch}c^2 = 4,52825 \cdot 10^{-5}$ eV in power units.

From the viewpoint of our approach, construction of the spectrum of the **ch**-particle through its overtones, similar to the electrino spectrum, sounds logical (see page 7). It should have an analogue of Balmer series where the first term is $hv^* = \frac{m_{ch}c^2}{2^2} = 9,0565 \cdot 10^{-5}$ eV, i.e. it exactly twice exceeds the energy of wave shift. In our understanding, the shift on the frequency, corresponding to $\Delta v$, takes place when the excitation energy exceeds the threshold corresponding to the electrino birth. Surplus of the brought energy is irradiated as quanta whose energy is a multiple energy of the rest mass of the **ch**-particle.

The value of super-thin splitting for hydrogen $\Delta v=1,420405 \cdot 10^9$ Hz ($\lambda$=0,211m, $hv$=5,87454·10⁻⁶ eV) was measured experimentally[11]. This value is obtained precisely from the ratio $hv = \frac{20}{3^2}\alpha m_{ch}c^2$. However, the author is at a loss to prove it physically.

Using ratios (5.9) it is possible to prolong the series of particles towards small energies. The next particle which we name a G-particle, should have a weight:

$$m_G c^2 = \frac{1}{2}\alpha^2 m_{ch}c^2 = 9,6454178 \cdot 10^{-9} eV \tag{5.12}$$

However, the experimental facts are needed to confirm the validity of such forecasts.

Thus, the critical points in which birth of particular leptons occur is designated on an energy scale. If to start from low energies, the rule (5.8) is observed up to electron. The hydrogen atom possesses enough internal energy for generation of younger leptons, and it reacts to weak external influence in a resonant manner. However, for generation of heavy leptons (muon and etc.) it is necessary to give extra energy to the proton. In this case the critical values are determined by the rule (5.2).

### 6. About magnetic moments

In the framework of the presented hypothesis, another idea arises on the magnetic moment of particles.

In the existing theory the magnetic moment **μ** is defined as a product of closed current **I** on streamline area **S**: **μ=IS**. Accepting that electron moves in atom on a circular orbit, a connection between the magnetic moment and the angular moment **L=h/2π=ℏ** is established.

$I = \dfrac{q}{T} = qv = \dfrac{q\omega}{2\pi}$     q – charge, **T** – orbital period, **v** – rotational frequency,

**ω** – angular velocity

**S=πr²**     **r** – orbital radius

$$\mu = \left(\dfrac{q\omega}{2\pi}\right) \cdot (\pi r^2) \cdot \dfrac{m}{m} = \dfrac{q}{2m} \cdot m(\omega r) r = \dfrac{q}{2m} mVr = \dfrac{q}{2m} L . \tag{6.1}$$

V – orbital velocity, L – angular moment.

Thus, the magnetic moment of the electron named as Bohr magneton is defined as $\mu_B = \dfrac{eh}{4\pi m_e} = 927.4 \cdot 10^{-26} JT^{-1}$. Accordingly, the magnetic moment of the proton should be ~1836 times less, i.e. $\mu_p = \dfrac{eh}{4\pi m_p} = 0.505078 \cdot 10^{-26} JT^{-1}$. But experiment gives other values: $(\mu_e)_{exp} = 928.476376 \cdot 10^{-26} JT^{-1}$, distinguished from $\mu_B$ by the factor of **a_e**=1.159652·10⁻³, named as "anomaly of the electron's magnetic moment"; $(\mu_p)_{exp} = 1.410606673 \cdot 10^{-26} JT^{-1}$. $\dfrac{\mu_p^{exp}}{\mu_p^{th}}$=2,79285. This value of the magnetic moment of the proton in nuclear magnetons is usually used in tabulated data. Such a significant divergence of experimental and theoretical values is explained by presence of virtual meson fields nearby the proton.

This way of calculating the magnetic moment of the proton, i.e. by dividing **μ_e** by the relation of masses of the proton and the electron, looks a little bit strange. In accordance with definition **μ=IS** it means that either the current created by the proton decreases 1836 times at the expense of reducing the frequency (the charge cannot decrease), or the area of the contour reduces the same time (radius decreases ~43 times). According to Bohr's model, the radius of the first electron's orbit is equal to 0,528·10⁻¹⁰m, accordingly the rotation radius of the proton should be 43 times less, i.e. about 1,23·10⁻¹²m, what is not real.

In the framework of the suggested model, the anomaly of the proton magnetic moment can be explained as follows. Considering the proton as a body undergoing simultaneously two rotations (the basic and precession ones), we have shown that the resonance arises, if the length of the wave corresponding to a particular periodic process is comparable to some geometrical sizes. Let us pay attention that the double Compton length of the proton $2\Lambda_p$ (accordingly, the 2 time smaller frequency, the overtone of the precession frequency) is equal to the length of the circle with radius **b** (a small half-axle of the ellipsoid-proton). **Λ_p**=1,32141·10⁻¹⁵m, $\mathbf{b}=1{,}48711 \cdot 10^{-16}$m. We also have established that the hit in the resonance occurs at a lower frequency (factor $\frac{10}{\pi^2}$). The condition of appearing such a resonance can be written as:

$$\frac{2\Lambda_p}{2\pi b} \cdot \frac{\pi^2}{10} = 2.7915 \qquad (6.3)$$

The obtained value differs from $\frac{\mu_p^{exp}}{\mu_p^{th}}$ by the factor of $\Delta=1{,}00047$. Thus, it is possible to assume that at experimental definition of the proton magnetic moment this resonance manifests itself at the frequency not 1836 times smaller than it was expected, but at the frequency $\frac{1836{,}1516}{2{,}7915} = 657{,}76$ times smaller the frequency of precession rotation. If to uphold the definition $\mu = \frac{q}{2m}L$, then the result that in agreement with experiment can be obtain as follows:

$$\mu_p = \frac{q}{2m_p}(m_p ca) \cdot \left(\frac{2\Lambda_p}{2\pi b} \cdot \frac{\pi^2}{10}\right) = 1{,}4099526 \cdot 10^{-26} JT^{-1}. \qquad (6.4)$$

In frames of the same definition at resonance corresponding $\frac{\Lambda_e}{\Lambda_p}$

$$\mu_e = \frac{q}{2m_p}(m_p ca) \cdot \frac{\Lambda_e}{\Lambda_p} = 927{,}4 \cdot 10^{-26} JT^{-1}. \qquad (6.5)$$

The obtained value coincides with Bohr`s magneton. In this case the factor $\frac{\pi^2}{10}$ should not be taken into account, because it has already been taken into account, when deriving (2.6). But in this formulation it turns out that Bohr`s magneton concerns not to electron but to proton in the condition of a resonance $\frac{\Lambda_e}{\Lambda_p}$. The anomaly of the magnetic moment observed in experiment $\mathbf{a_e}=1{,}1596 \cdot 10^{-3}$ comes to light as follows. Return to the formula (5.4), according to which $\frac{(LV)_{sp}}{hc} = \frac{1}{8\pi^4\left(1+\frac{1}{2\pi^2}\right)^2} = 1{,}1624 \cdot 10^{-3}$. It practically coincides with value $\mathbf{a_e}$. Entering into (6.5) the correction taking into account the spin rotation, we obtain

$$\mu_e = \frac{q}{2m_p}(m_p ca) \cdot \frac{\Lambda_e}{\Lambda_p}\left(1+\frac{(LV)_{sp}}{hc}\right) = 928{,}478 \cdot 10^{-26} JT^{-1}, \qquad (6.5a)$$

i.e. the value that precisely coincides with the experimental value. Introduction of this correction means that in the condition of resonance $\frac{\Lambda_e}{\Lambda_p}$ the proton has an additional precession and, accordingly, an additional rotation speed.

Similar way the neutron's magnetic moment can be agreed with resonance $\frac{2\Lambda_p}{2\pi a}$. Thus,

$$\mu_n = \frac{q}{2m_n}(m_p ca) \cdot \left(\frac{2\Lambda_p}{2\pi a} \cdot \frac{\pi^2}{10}\right) = 0{,}99559 \cdot 10^{-26} JT^{-1} \qquad (6.6)$$

Discrepancy with $(\mathbf{\mu_n})_{exp}=-0{,}966236 \cdot 10^{-26} JT^{-1}$ is 3%. The reasons of such a divergence are a subject to be discussed.

Thus, in the suggested model the manifestation of the magnetic moment is reduced to a resonant effect. It is in agreement with the experimental practice in which the resonant methods of measurement of the magnetic moments are used.

### 7. X-ray spectra and a nucleus

*X-ray spectra*

To show generality of the selected approach in the definition of the radiation spectra, we apply it to definition of spectra of characteristic X-ray radiation. Research on these spectra has shown that with reference to a specific chemical element with atomic number Z the structure of the spectrum submits the same ratio as the structure of optical spectra of a hydrogen atom. For each element some series are allocated: K, L, M …, in which the observable wavelengths are determined by relation:

$$\frac{1}{\lambda_{K\alpha}} = R(Z-1)^2 \left(\frac{1}{1^2} - \frac{1}{2^2}\right) \qquad \frac{1}{\lambda_{L\alpha}} = R(Z-\sigma)^2 \left(\frac{1}{2^2} - \frac{1}{3^2}\right) \qquad (7.1)$$

where $\lambda$ – the wavelength in a corresponding series, **R**=1.0973742·10⁷m⁻¹ – Rydberg constant, **σ** - some constant number, different for series L, M … The dependence of wave lengths $\lambda_{K\acute{a}}$ and $\lambda_{L\acute{a}}$ from Z is shown on figs.2,3. For convenience some values are presented on fig. 4,5 in power units (in accordance with $\lambda = \frac{hc}{h\nu}$).

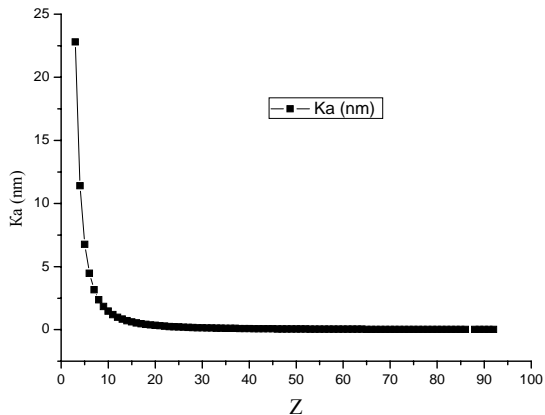

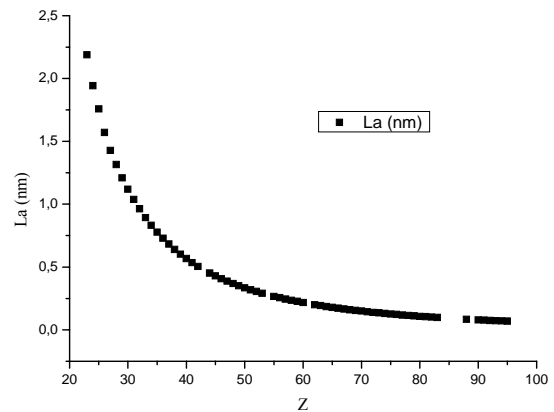

Fig.2 The wave length $\lambda_{K\acute{a}}$=f(Z)   Fig.3 The wave length $\lambda_{L\acute{a}}$=f(Z)

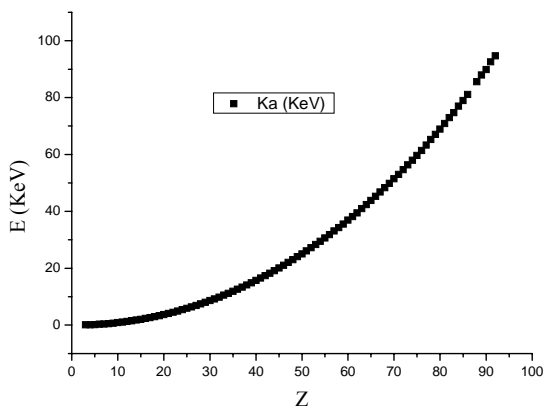

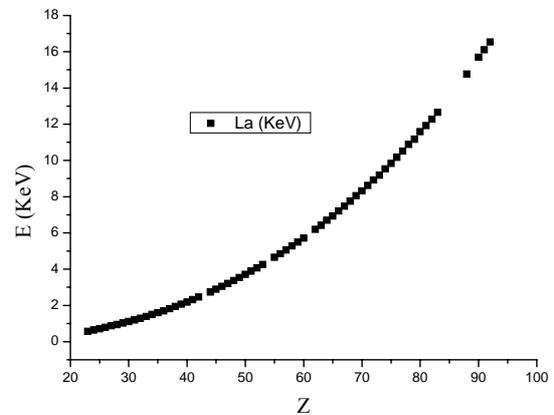

Fig.4 Energy of quantums $E_{K\acute{a}}$=f(Z)   Fig.5 Energy of quantums $E_{L\acute{a}}$=f(Z)

The form of expressions (7.1) is similar to expressions for definition of the structure of series of an optical spectrum of a hydrogen atom. First is similar to definition of Layman series and has a difference that **R** is replaced **R(Z-1)²**. The second expression is similar to definition of Balmer series; accordingly, **R** is replaced **R(Z-σ)²**. Such a replacement follows from the empirical Mozely law that has found out a connection between values $\lambda_{K\alpha}$ and **Z** (7.2).

$$Q_{K_\alpha} = \sqrt{\frac{1}{0.75 \lambda_{K\alpha} R}} \simeq Z - 1 \qquad (7.2)$$

Practically the linear dependence of number $Q_{K\alpha}$ from **Z** (fig. 6) has initiated a lot of hypotheses about the nature of the characteristic X-ray radiation. They were reduced to the following: a) this radiation arises at energy transitions in internal shells of atoms, b) electrons rotating around the nucleus shield in part a nucleus charge, c) Rydberg constant plays a particular fundamental role. It is necessary to note that Mozely law is of approximated character. The error at the determination of Z is shown on fig. 7.

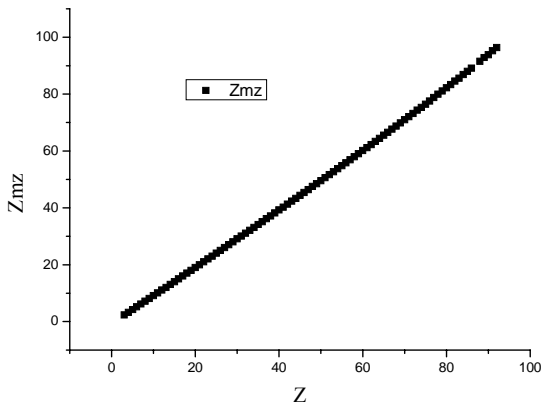
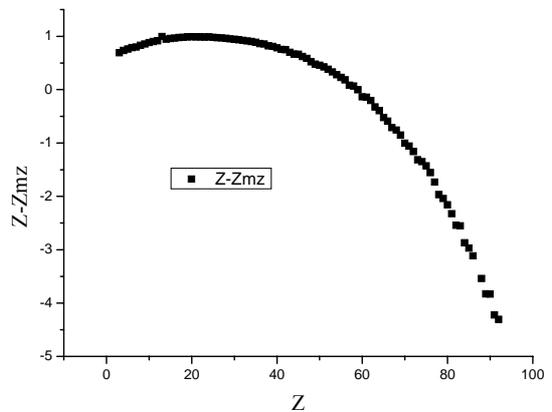

Fig.6 The Mozely law    Fig.7 The error at the determination of Z on Mozely law

In the suggested model the other interrelation between the parameters of the nucleus is established: atomic number Z, mass number M and in the characteristic X-ray wavelengths of its characteristic X-ray $\lambda_{K\acute{a}}$ и $\lambda_{L\acute{a}}$.

We present experimental data on K and L-series as dependences $\frac{E_{K\alpha, L\alpha}}{Z} = f(Z)$. Figs.8,9 show that these functions are of approximately linear character. The relation of the first members of the series is equal to 5,4 and close to the energy relation of the first members of Layman and Balmer series (see tab. 1,2).

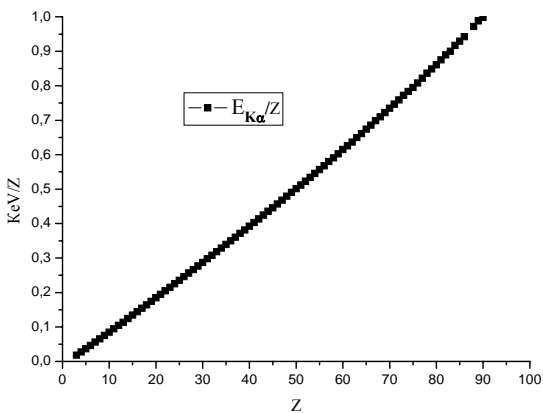
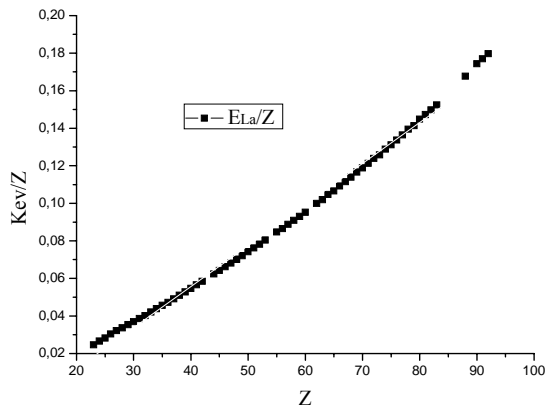

Fig.8  $E_{K\acute{a}}/Z = f(Z)$    Fig.9  $E_{L\acute{a}}/Z = f(Z)$

Let us show how the mass of nucleus M influence energy $E_{K\alpha}$ and how the interrelation between $E_{K\alpha}$, M and Z is established in frames of the suggested model. Fig.10 shows dependence $E_{K\alpha}$=f(ZM$_{mono}$), where M$_{mono}$ - mass of monoisotopes in standard a.м.u. (further we shall use these units for M).

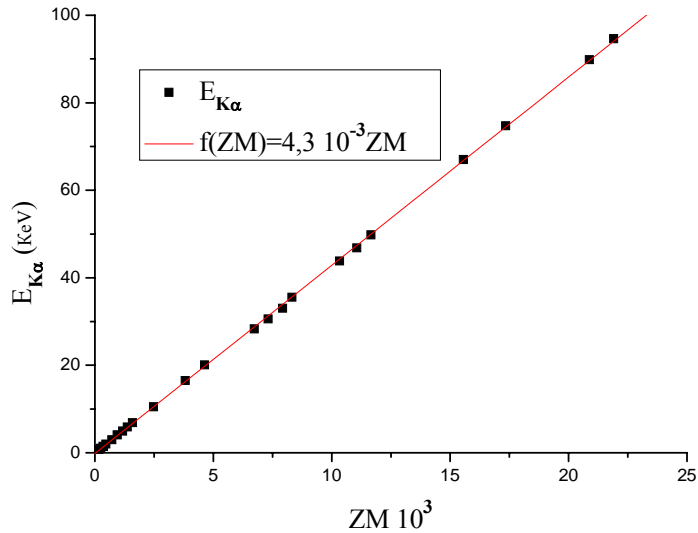

Fig.10 $E_{K\alpha}$ = f(ZM)

One can see that this function is linear:
$$\frac{E_{K\alpha}}{ZM} = \varepsilon \approx const, \qquad (7.3)$$
where $E_{K\alpha}$ (KeV), Z - atomic number of the element, M - mass of nucleus (a.м.u.), $\varepsilon = 4{,}3 \cdot 10^{-3}$ KeV/zM. The calculation of value $E_{K\alpha}$ coincides with experiment within 1.0 % (see Tabl.3).

Table 3

The experimental and calculated data on $E_{K\alpha}$

| Isotope | $E_{K\alpha}$ exp. KeV | $E_{K\alpha}$ calc. KeV | Δ KeV |
|---|---|---|---|
| $_{92}U^{238}$ | 94.656 | 94.153 | 0.503 |
| $_{69}Tm^{169}$ | 49.799 | 50.142 | 0.343 |
| $_{45}Rh^{103}$ | 20.065 | 19.930 | 0.135 |
| $_{27}Co^{59}$ | 6.916 | 6.850 | 0.066 |

As K-series and optical Layman series have an identical structure, submitting to the rule $\frac{1}{\lambda} = const\left(\frac{1}{1^2} - \frac{1}{2^2}\right)$, we can assume that the K-series is a continuation of Layman series for atoms with big Z.

Let us find out the ratio of the energies of K-series lines $E_{K\alpha}$=f(Z) and the ionization energies of hydrogen-like atoms $E_{ion}^{\sim H+}$=f(Z), i.e. completely ionized atoms: $He^{2+}$, $Li^{3+}$, $Be^{4+}$, etc. up to $Kr^{36+}$. It is known that the structure of optical spectra of hydrogen-like alkaline atoms (group 1A Mendeleyev's tables) is identical. The energy $E_{ion}^{\sim H+}$=$Z^2 \cdot E_{ion}^{H+}$=13,605667$Z^2$ eV. This formula is held precisely enough. The relation of the calculated and experimental values is

$$\frac{\left(E_{ion}^{\sim H+}\right)_{cal}}{\left(E_{ion}^{\sim H+}\right)_{exp}} = 1 - 4{,}8 \cdot 10^{-5} Z + 1{,}434 \cdot 10^{-5} Z^2.$$ For example, for $Kr^{36+}$ it is equal to 1,017. The forms of curves $E_{ion}^{\sim H+} = f(Z)$ and $\mathbf{E_{K\acute{a}}} = f(Z)$ are identical. Dependences $E_{ion}^{\sim H+} = 6{,}11 \cdot 10^{-3} ZM$ and $\mathbf{E_{K\acute{a}}} = 4{,}3 \cdot 10^{-3} ZM$ also have a linear character (fig. 11 and fig. 10). Inclinations of these straight lines differ. Similar curves take place for ions with smaller $Z_{ion}$.

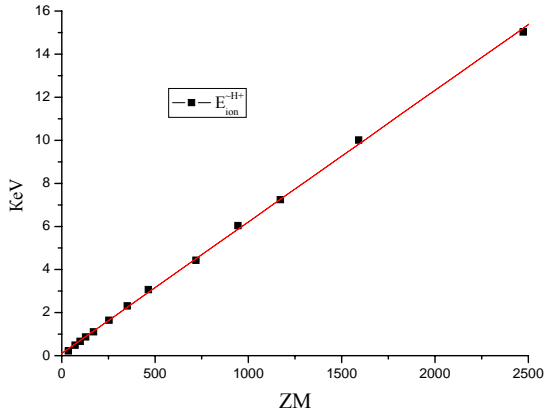
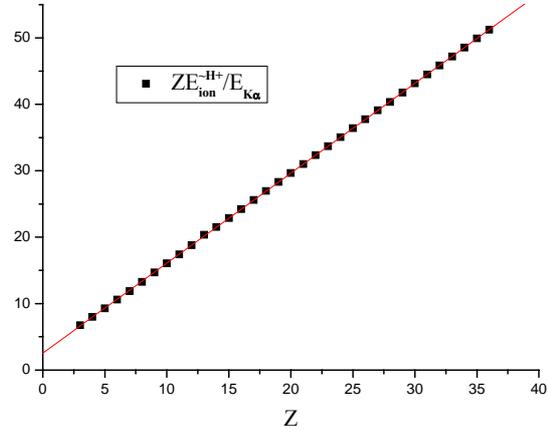

Fig.11 $E_{ion}^{\sim H+} = f(ZM)$      Fig.12 $Z \dfrac{E_{ion}^{\sim H+}}{E_{K\alpha}} = f(Z)$ in KeV/ KeV

However, we shall pay attention only to dependence $\dfrac{E_{ion}^{\sim H+}}{E_{K\alpha}} = f(Z)$ in range Z=1-36. Fig.12 gives the expression $Z \dfrac{E_{ion}^{\sim H+}}{E_{K\alpha}} = f(Z)$ which is an accurate straight line. It is described by empirical expression f(Z) =2,56249+1,3581Z. Extrapolating it to $H^+$, we obtain $\dfrac{E_{ion}^{H+}}{E_{K\alpha}} = 3{.}9206$, or at $E_{ion}^{H+} = 13{,}605667$ eV (the value determined on an optical limit) $\mathbf{E_{K\acute{a}}} = 3{,}469$ eV, i.e. the value coincident with $\mathbf{h\nu^*}$ at $\mathbf{n=2}$ for Layman series with accuracy of 2 % (see Tab.1). Thus, Layman series and K-series sew together on hydrogen.

*Masses of nuclei*

The revealed dependence $\mathbf{E_{K\acute{a}}} = f(ZM)$ provides away for determining the mass M at known Z and $\mathbf{E_{K\acute{a}}}$. However, the accuracy of M calculation under expression (7.3) is acceptable but not sufficient. The matter is that the coefficient $\varepsilon$ is not strictly a constant. It is shown on fig.13. Fig.14 shows a difference $M_{cal} - M_{exp} = \Delta M = f(Z)$ where $M_{cal}$ is calculated from (7.3): $M_{cal} = \dfrac{E_{K\alpha}^{exp}}{4{,}3 \cdot 10^{-3} Z}$. It is seen that $\Delta M$ has a wide scatter up to 6 a.m.u. The reason of this disorder will be found out below.

It is visible that dependence $\varepsilon = f(Z)$ has a similarity with the dependence $\dfrac{M}{Z} = f(M_{exp})$ shown on fig.15. Besides the masses of stable nuclei, the masses of the most long-living isotopes are included in the area Z> 92: $_{93}Np^{237}$, $_{94}Pu^{244}$, $_{95}Am^{243}$, $_{96}Cm^{247}$, $_{97}Bk^{247}$, $_{98}Cf^{251}$, $_{99}Es^{254}$, $_{100}Fm^{257}$, $_{101}Md^{258}$, $_{102}Nb^{255}$, $_{103}Lr^{262}$, $_{104}Rf^{261}$, $_{105}Db^{262}$, $_{106}Sg^{266}$, $_{107}Bh^{264}$. It is possible to see correlations in wavy behaviour of curves on fig. 14 and fig. 15. As a whole, the curve $\frac{M}{Z} = f(M_{exp})$ has a smooth growth with small recessions aside big M. It is seen that these recessions coincide with the advent of natural radio-activity. They are marked on fig. 15. Restoration of stability takes places quite quickly. For example, in the area behind Bi (Po-Ra) it needs 5 atomic numbers. This allows one to make a conclusion that restoration of the relative stability will take place and in the area Z>107 much earlier than it is supposed according to theoretical representations available now. Identification of isotope $_{105}Db^{268}$ with $T_{1/2}$=25h in [12] is the indication on the competency of such an assumption. Fig.15 gives a hypothetical stable nucleus $_{114}A^{298}$. It is seen that curve $\frac{M}{Z} = f(M_{exp})$ tends to this point.

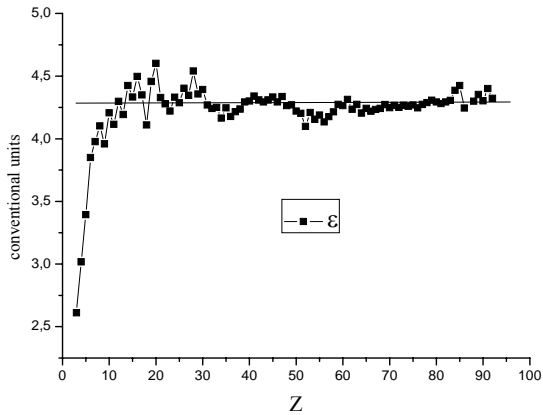

Fig.13 Dependence ε=f(Z)

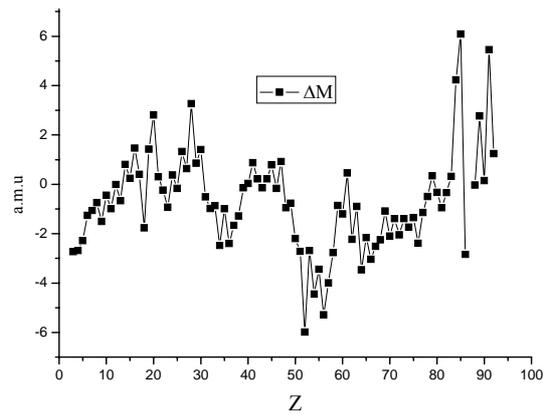

Fig.14 $M_{exp} - M_{cal} = \Delta M = f(Z)$

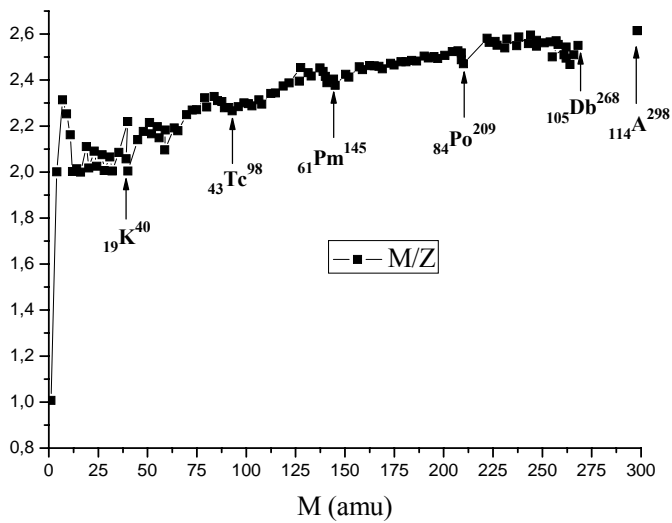

Fig.15 $\frac{M_{exp}}{Z} = f(M_{exp})$

Dependence $\frac{M_{exp}}{Z} = f(Z)$ is of irregular character (fig.16). But dependence $\frac{M}{Z} + \varepsilon Z = f(Z)$ at ε=4,318 is a definitely straight line f(Z)=2+4,32454Z, and we can write the empirical formula:

$$\frac{M}{Z} = 2 + 6,4204 \cdot 10^{-3} Z \qquad (7.4)$$

In essence, it is the line of stability (fig.17) specifying the number of coupled protons and neutrons + a neutron addition. The physical sense of the formula (7.4) is not clear, but the result is obvious.

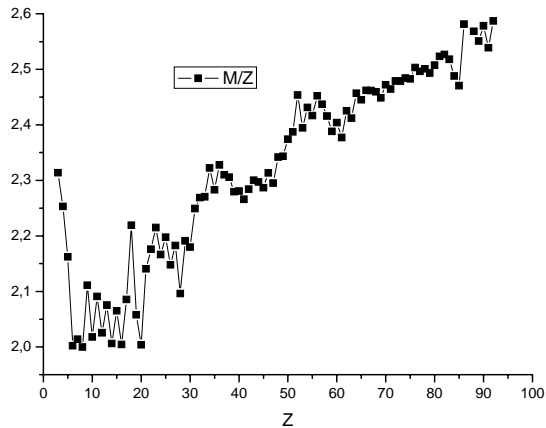 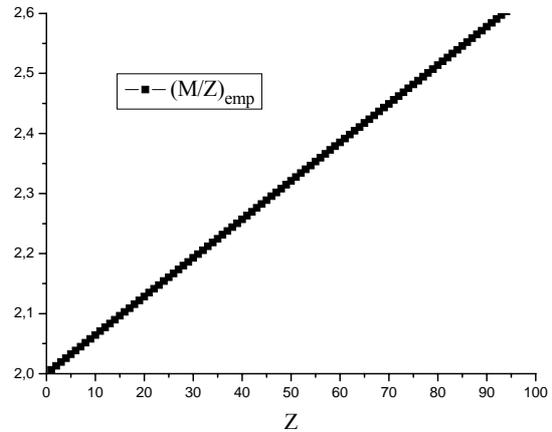

Fig.16 $\frac{M}{Z} = f(Z)$    Fig.17 $\frac{M}{Z} = 2 + 6,4204 \cdot 10^{-3} Z$

The formula (7.4) allows one to calculate nuclei masses at Z <92 quite precisely. For example, for monoisotopes $M_U$=238,34 a.m.u., $M_{Th}$=232,00 a.m.u., $M_{Rh}$=103,00 a.m.u., $M_{Co}$=58,68 a.m.u.. But not everything is so good. Fig.18 shows deviations of value $\frac{M_{cal}}{M_{exp}}$ from 1. Let us analyze them.

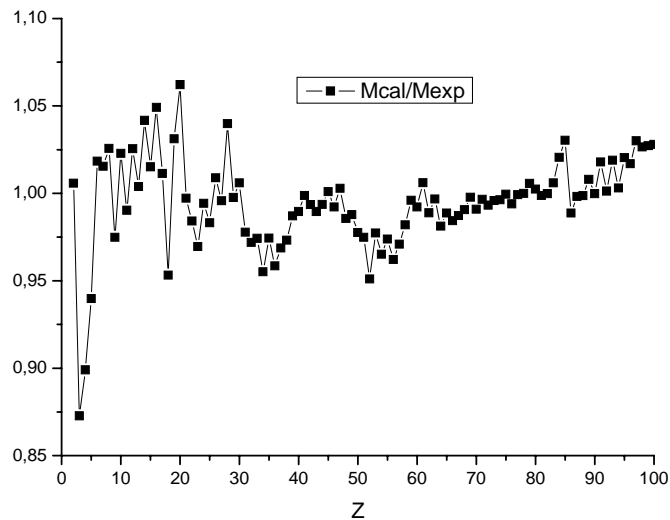

Fig.18 $M_{cal}/M_{exp}=f(Z)$

Let us show that deviations $\frac{M_{cal}}{M_{exp}}$ are connected to the periodicity of density of elements $\rho$=f(Z) and, accordingly, of the atomic volume. Fig.19 gives dependence $\rho$=f(Z), it is of periodic character. Fig.20 presents a dependence of nuclear volume $V_{at}$=f(Z), obtained by division $M_{exp}$ by $\rho$. It follows Mendeleyev's Periodic Law with allocation of peaks of alkaline elements. On the average $V_{at}$~16·10$^{-30}$m$^3$; accordingly, is the characteristic size of atom $a$~2,5·10$^{-10}$m. Dependence $V_{at}$=f(Z) has been established by Meyer at the end of the XIX century. It is considered that $V_{at}$ does not mean the volume of one atom, but the volume of many atoms in a volume unit. It was also considered that the periodicity in the properties of elements is in no way reflected on the dependence $E_{Kά}$=f(Z) as X-ray spectra arise inside the atom, nearby a nucleus, instead of the periphery of the atom where optical spectra appear and where the action of the nucleus charge completely weakens due to shielding of the internal electrons cloud [4]. Now there is reason to be doubtful of this statement.

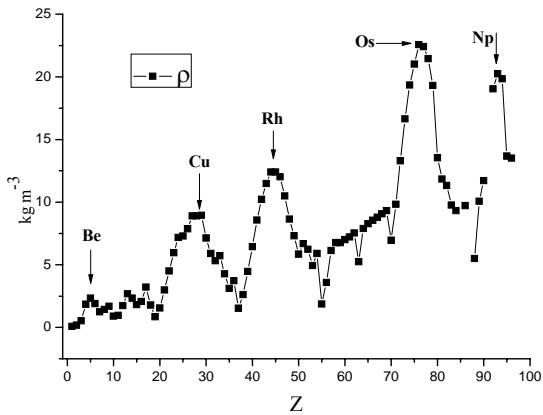 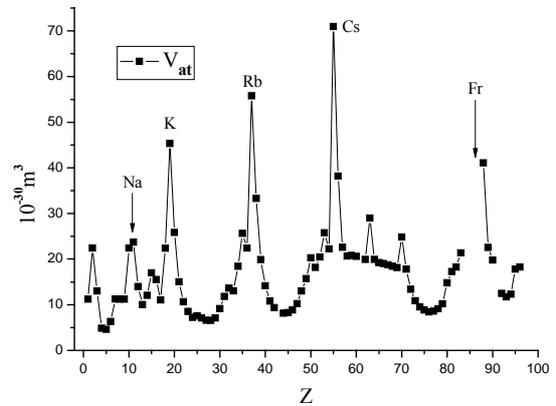

Fig.19 Density of elements **ρ** = f(Z)         Fig.20 Atomic volume $V_{at}$ = f(Z)

Fig. 21 shows the dependence $\frac{M_{cal} - M_{exp}}{\rho} = \frac{\Delta M}{\rho} = \Delta V_{at} = f(Z)$ where $M_{cal}$ is calculated from (7.4). It is seen that the error in the definition of mass ΔM correlates with nuclear volume $V_{at}$ and varies according to $\Delta V_{at}$=f(Z). Taking into consideration that **ρ**=f(Z) as a whole is proportional to Z at all fluctuations, we shall construct dependence $\frac{\Delta M}{\rho Z} = f(Z)$ (fig. 22). One can see that it varies rather poorly, and in the area Z> 60 $\frac{\Delta M}{\rho Z} \approx 0$. If we knew analytically the law **ρ**=f(Z), we would calculate precisely the weights of nuclei.

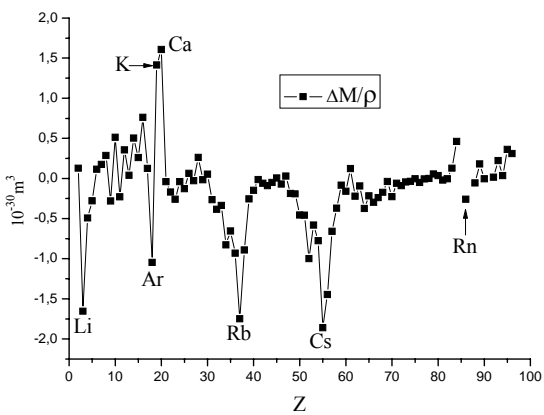 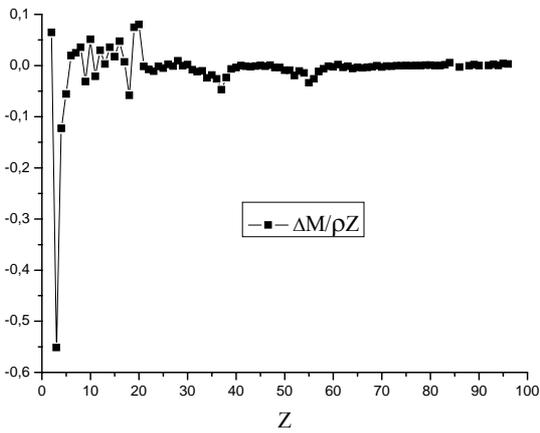

Fig.21 $\frac{\Delta M}{\rho} = \Delta V_{at} = f(Z)$         Fig.22 $\frac{\Delta M}{\rho Z} = f(Z)$

It is necessary to notice that we seemingly to run into contradiction by accepting the mass distributed on the whole volume of the atom but not concentrated in its nucleus. As we have found out earlier, the sizes of the proton are: $\mathbf{b} \approx 1.5 \cdot 10^{-16}$m and $\mathbf{a} = \sqrt{2}\,\mathbf{b} \approx 2.1 \cdot 10^{-16}$m, accordingly $V_p = \frac{4}{3}\pi ab^2 \approx 20 \cdot 10^{-48}$m³ (ellipsoid volume). In accordance with fig. 20, the volume of hydrogen atom is equal to $V_H \approx 10 \cdot 10^{-30}$m³ и ř=1,3·10⁻¹⁰м (volume and radius of a sphere).

Therefore one can conclude that the density of matter in the atom changes 6 orders along its linear size from the nucleus to periphery. But under what law?

### *About the chemical nature of elements*

The further analysis of the known experimental data shows a number of interesting peculiarities in the periodicity of the atom properties. They are related to the periodicity of nuclear volume $V_{at}$=f(Z) and to the relation $\frac{M}{Z}$ as well.

It is known that the energy of single ionization of atoms $E_{ion}$=f(Z) correlates with $V_{at}$=f(Z). If to compare the dependence $E_{ion}$=f(Z) with $\frac{M}{Z}$ as a ratio (7.5)

$$E_{ion} = \frac{Z}{\chi M_{exp}}, \qquad (7.5)$$

then we get an interesting picture It is presented on fig. 23 as dependence $\chi$=f(Z), where $\chi$ - the factor connecting $E_{ion}$, Z and $M_{exp}$ of a particular atom. It is named as "factor of individuality".

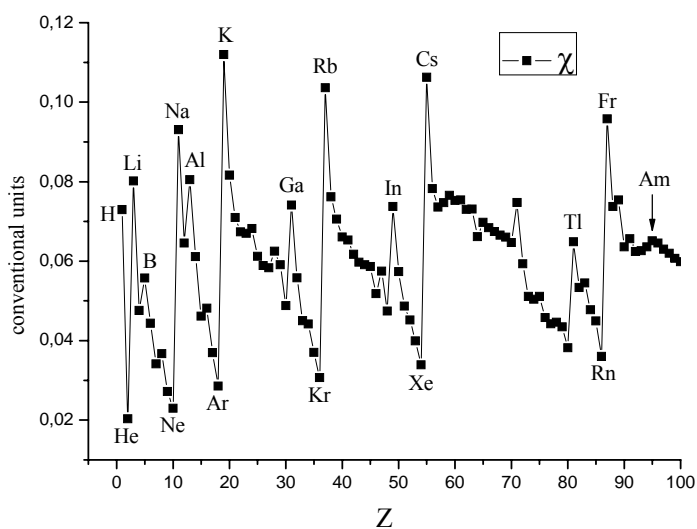

Fig.23 Factor of individuality $\chi$=f(Z)

This figure is a surprisingly informative one. The periodicity of the chemical properties of elements obtains its certain quantitative expression. As a whole, the picture is similar to a unwrapping accordion. The periods of Mendeleyev`s table (from alkaline elements up to noble gases) are as though stretched keeping their similarity. This is clearly seen on elements B, Al, Ga, In, Tl incorporated in column 3A of the Periodic table. It is known that the number of elements N in the periods submits rule N=2n$^2$, **n** - an integer, i.e. N=2, 8, 18, 32 … Numbers 8 and 18 repeat twice: $_{10}$Ne – $_2$He, $_{18}$Ar – $_{10}$Ne и $_{36}$Kr – $_{18}$Ar, $_{54}$Xe – $_{36}$Kr, number 32 "works" from $_{54}$Xe up to $_{86}$Rn. The behaviour of dependence $\chi$ =f (Z) behind Fr suggests that the next period can contain more than 32 elements. Naturally, the dependence $\chi$ =f (Z) should be discussed, but it is not a subject of this work.

### *General remarks*

The mentioned representation of experimental data on X-ray radiation gives basis to draw some general conclusions. The proton and the atoms of elements are as a whole the periodic system resounding on an external influence. Reaction to external influence in the process of increasing its energy is radiation of particles in the critical points determined in (2.15), while in the intervals - electromagnetic radiation, the spectrum of which submits to harmonious rules. The atoms of each element are individual on their physical and chemical properties. It is generally agreed that this individuality is defined, first of all, by atomic number Z, namely by the number of protons in a nucleus. In the suggested concept the role of Z looks different.

We have convinced that the reaction of atom to external influence in an X-ray energy range is expressed by X-ray characteristic radiation (or radiation of Auger electrons). We have convinced that the energy of X-ray quanta depends not only on Z but it is directly proportional to the product ZM with small variations of coefficient ε, depending upon value $\frac{M}{Z}$ and upon change of the atomic volume. If to take that the nucleus of an atom, similar to a proton, is of ellipsoid form and is in a free precession, then it looks important to know the energy of the main rotation and its velocity after the nucleus react to external influence by changing the basic rotation velocity in a resonant manner. According to (2.2) $E_{K\alpha} = 0,1 M V_{main}^2$ and according to (7.3) $E_{K\alpha} = \varepsilon Z M$. Having equated them, we obtain $\frac{0,1 V_{main}^2}{\varepsilon} = Z^*$, i.e. Z in conditional units $\frac{V_{main}^2}{\varepsilon}$. The dependence $V_{main}^2 = f(Z)$, calculated from $V_{main}^2 = \frac{E_{K\alpha}}{0,1M}$ is presented on fig.24. Dividing it by 10ε, we obtain $V_{main}^2$ in terms of Z. This dependence is presented on fig.25. It is a straight line Z*=9,649·10⁻²Z, and we can conclude that we get a direct proportionality between $V_{main}^2$ and Z.

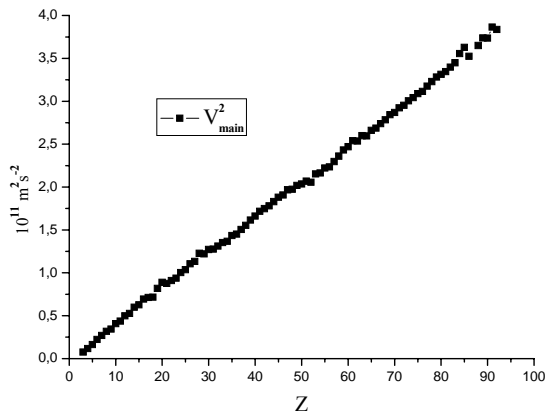 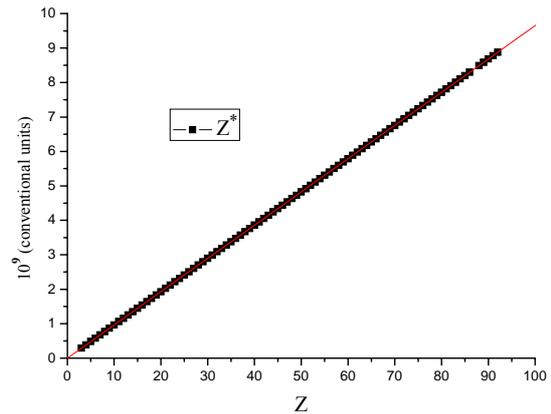

Fig.24 $V_{main}^2 = f(Z)$      Fig.25 Z*=9,64853·10⁻⁴Z

Why does the number of protons Z in a nucleus only set the speed of rotation? An explanation to this fact is that number Z for each particular nucleus designates a resonance arising at external influence on the nucleus as a whole rotating body. These resonances are distributed in a quadratic manner over the power scale and linearly over scale ZM. But it does not mean that the nucleus is a continuous body. It consists of separate parts which we name protons and neutrons forming a granular structure. They, in turn, undergo rotation connecting with each other someway and forming stable structures of various configurations. Deviations from the stable structure under effect of external influence or internal movement result in radioactive processes.

By what is the stability determined? This is a special question. In paper[13] there is an attempt to find some general regularities. The author proceeds from an abstract principle, "the whole is stable, if it consists of the greatest number of parts, each is different one from another". He has deduced mathematical expressions for calculation of magic numbers A and atomic numbers C for noble gases (see 7.6a,b). The number A means the maximal number of rotary degrees of freedom (including both the right-hand and left-hand rotation) realized in the system of identical particles forming a shell, plus 2 rotary degrees of freedom of the system as the whole. Numbers A and C are formed in essence by a uniform way with use of tools of combinatory mathematics. The numbers of natural scale **n=1**, 2, 3 … are some quantum numbers. The known magic numbers 8, 20 and 40 drop out from the obtained row, but the property of additivity of magic numbers is known too. Therefore, the missing numbers are formed as 2+6=8, 6+14=20, 20+20=40. The numbers C is formed by deduction from A of member $2n^2$ corresponding to the numbers of filling the periods of the Periodic system. It can be named as a figure game, but it is possible that such an approach reflects physical contents. Formally the formulas (7.6a,b) are similar to quantum-mechanical expressions for description of the structure of atoms and nuclei. Probably, it is possible to find decisions of specific tasks on this way.

$$A = \sum_{m=0}^{n}\left[m(m+1)+2\right] = \frac{(n+1)^3 + 5(n+1)}{3} \qquad (7.6a)$$

| n | 0 | 1 | 2  | 3  | 4  | 5  | 6   | 7   |
|---|---|---|----|----|----|----|-----|-----|
| A | 2 | 6 | 14 | 28 | 50 | 82 | 126 | 184 |

$$C = \sum_{m=0}^{n}\left[(m-1)(m+2)\right] = \frac{(n+1)^3 + 5(n+1)}{3} - 2n^2 \qquad (7.6б)$$

| n | 0 | 1 | 2 | 3  | 4  | 5  | 6  | 7  |
|---|---|---|---|----|----|----|----|----|
| C | 2 | 4 | 6 | 10 | 18 | 32 | 54 | 86 |

### 7. Conclusion

From all the variety of physical processes we have allocated the particles related to weak interactions. Is it possible to apply the suggested approach to strong interactions? The author did not study this question specially. However, Gareeev's paper[14] gives a new systematization of elementary particle resonances using a resonant approach. Practically all masses of resonances have been calculated with surprising accuracy in the assumption that its decay impulses are multiple to a decay impulse of a pion: **$P_n=nP_\pi$**; **n**=1,2,3…, **$P_\pi$**=29,7918 MeV/c.

The idea of similarity of the physical processes available in various energy intervals is not a new one. V.Weisskopf[15] stated the generalizing ideas on the structure of atomic optical spectra, nuclear spectra and a mass spectrum of elementary particles. R.Feynman[16] built physical processes along a frequency scale designating critical points. In the presented paper such critical points are designated by birth of leptons with a certain mass. The author did not put the task to analyze in details the processes in each energy range and to review the existing theoretical representations. In the present paper a different original interpretation of the known experimental facts is given at estimation level only.

The hypothesis suggested in this paper is based on the principles of classical physics. It contains a completely different sight at the nature of elementary physical processes. The simplicity and accuracy of the found decisions allow one to approve that the hypothesis is correct in its basis. At the same time, it is paradoxical. It is difficult to imagine that the free precession is the basic state of the proton. A necessary condition is a predetermined form of the proton as an ellipsoid (egg). Another necessary condition is the equality of the velocity of precession to the velocity of light, what is in a drastic contradiction with the available representations. The mass of particles and electromagnetic radiation are different concepts in the idea, as distinct from the supposed electromagnetic origin of mass. At the same time, for birth of heavy leptons as individual objects with a fixed mass it is required to spend energy. What way the energy converts into the mass is an ancient question, but the given hypothesis does not make it more clear. The refusal of the concepts of electromagnetism has not prevented us to determine correctly the structure of optical spectra. In addition, the concepts of electric charge and magnetic moments have received a mechanistic interpretation. Definition of spin energy has enabled to connect weights of leptons through a constant of thin structure. In the hypothesis the quantum nature of nuclear processes has the other content and reduces to resonant processes of periodic systems. The author realizes perfectly well that such serious statements look irritating, but he hopes that physicists will apprehend the presented results seriously.